# Abatement of Computational Issues Associated with Dark Modes in Optical Metamaterials

Matthew LePain
Georgia Southern University
Physics Department
ml03213@georgiasouthern.edu

Maxim Durach
Georgia Southern University
Physics Department
mdurach@georgiasouthern.edu

## ABSTRACT
Optical fields in metamaterial nanostructures can be separated into bright modes, whose dispersion is typically described by effective medium parameters, and dark fluctuating fields. Such combination of propagating and evanescent modes poses a serious numerical complication due to poorly conditioned systems of equations for the amplitudes of the modes. We propose a numerical scheme based on a transfer matrix approach, which resolves this issue for a parallel plate metal-dielectric metamaterial, and demonstrate its effectiveness.

**Categories and Subject Descriptors**
J.2 [**Physical Sciences And Engineering**]: Physics.

**General Terms**
Nanotechnology.

**Keywords**
Photonics, Plasmonics, Metasurfaces.

## 1. INTRODUCTION
Modern nanotechnology poses a plethora of cutting-edge research problems in the fields of nano-optics and electronics, which are ideal for reinforcement of the knowledge gained in the upper division physics courses, such as Classical Electromagnetics and Quantum Mechanics, as well as for training in numerical methods and computational techniques. Due to exquisite spatial profile of nanostructures, the solutions to these problems feature combinations of propagating and evanescent waves. This is known to pose considerable numerical complications if care is not taken. In particular, applying the straightforward routine of setting boundary conditions at nanostructure boundaries results in poorly conditioned systems of equations and unacceptable errors due to evanescent waves. This has been discussed for a number of optical structures, including stratified media [1], and sine-wave grating [2].

Plasmonic metamaterials and metasurfaces is a rapidly developing field, which encompasses such phenomena as negative refraction [3], superlensing [4], optical cloaking [5], wavefront control [6] and much more. Plasmons are evanescent waves bound to the interfaces between metal and dielectric materials. The new functionalities are achieved when metal-dielectric structures feature subwavelength design forming metamaterials. The bright modes of these structures behave according to the effective metamaterial medium approximation, whereas the dark plasmonic modes are strongly localized. This leads to the numerical issues related to presence of both propagating and evanescent fields to be strongly expressed in metamaterial structures.

## 2. PROBLEM FORMULATION
In this paper, we present a comparison of two techniques to calculate electromagnetic fields in a nanostructure, which contains an array of nanoscale metal plates separated by layers of high-index dielectric placed above a transparent substrate. This problem is very important for the fields of photonics and metamaterials and its solution will allow the modeling of ultra-thin polarization rotators and nanoscale light emitters with controlled polarization [7].

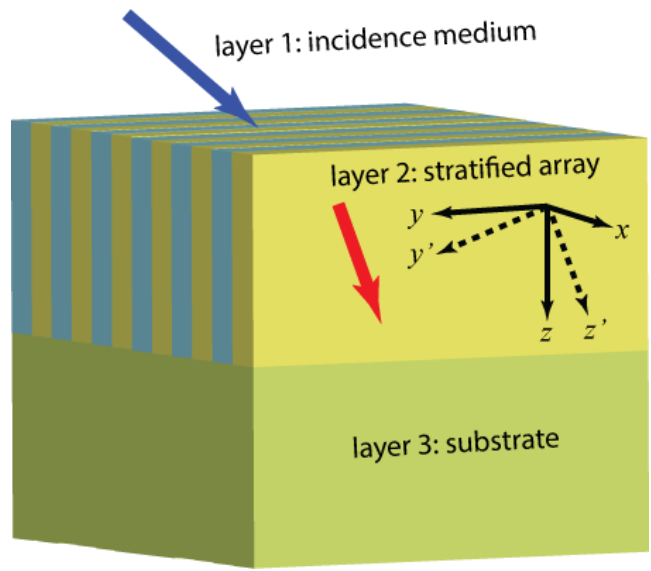

**Figure 1: Structure Schematics. The three-layer structure considered in this paper contains a one-dimensional metamaterial in layer 2. Note that the selection of coordinates that are shown here are explained in the text.**

From the computational perspective this structure requires simultaneous consideration of propagating and evanescent waves, therefore the boundary condition equations are numerically unstable [1, 2]. We devise a transfer matrix computational technique specific to this structure to resolve this issue by dynamically removing the evanescent waves from the computation as they decay in the structure.

## 3. METHODS

Consider a three-layered structure composed of layers 1 and 3, which are homogeneous and isotropic and layer 2, which is a one dimensionally periodic array of two different homogeneous and isotropic materials (Fig. 1). Because of the periodicity of layer 2 diffraction waves will be produced in layers 1 and 3 with diffraction wave vectors

$$k_x^{(n)} = k_x + \frac{2\pi n}{d} \quad (1)$$

Here $d$ is the period of the structure. The fields in layer 1 will be represented as:

$$\boldsymbol{F} = Re\left[e^{i\omega t}\left(I\, e^{i\boldsymbol{k_0}\cdot\boldsymbol{r}} + \sum_{n=-\infty}^{\infty} R_n\, e^{-i\boldsymbol{k_n}\cdot\boldsymbol{r}}\right)\right]\hat{\boldsymbol{f}} \quad (2)$$

where $F$ is either the magnetic field $H$ (for TM polarization) or the electric field $E$ (TE polarization) and $\hat{\boldsymbol{f}}$ is in the transverse direction. The $R_n$s are amplitudes of the reflected waves, and $I = P$ for TM fields and $I = S$ for TE fields is the incident wave amplitude. Also $\omega$ is the angular frequency, $\boldsymbol{k_0} = k_x\hat{\boldsymbol{x}} + k_y\hat{\boldsymbol{y}} + k_z\hat{\boldsymbol{z}}$, $\boldsymbol{r}$ is the position vector, and $\boldsymbol{k_n} = k_x^{(n)}\hat{\boldsymbol{x}} + k_y\hat{\boldsymbol{y}} + \sqrt{k_0^2\varepsilon_I - k_x^{(n)2} - k_y^2}\hat{\boldsymbol{z}}$. In layer 3 the fields are represented as:

$$\boldsymbol{F} = Re\left[e^{i\omega t}\sum_{n=-\infty}^{\infty} T_n\, e^{i\boldsymbol{k_n}\cdot\boldsymbol{r}}\right]\hat{\boldsymbol{f}} \quad (3)$$

Where the $T_n$s are amplitudes of transmitted waves.

The fields in layer 2 are more complicated. Because of the reflections on the layers' boundaries, waves that propagate in the positive and negative *x* directions are present in each material. In the case that the incidence plane is at an angle to the stratification of layer 2 (*x* direction), the convenient directions in which to define polarization are different within each layer. This leads to the fields being excessively complicated to solve in the *x-y-z* coordinate system. Thus, we simply consider a wave propagating in the $z'$ direction and rotate the coordinates back when convenient (see Fig. 1). The un-rotated field equations look like this:

$$F_{y'}^{(m)} = Re\left[e^{i\omega t}\left(A_m e^{ik_{z'}^{(m)}z'} + B_m e^{-ik_{z'}^{(m)}z'}\right) \times \begin{cases} C_m e^{i\alpha_1 x} + D_m e^{-i\alpha_1 x} & \text{Material 1} \\ G_m e^{i\alpha_2 x} + J_m e^{-i\alpha_2 x} & \text{Material 2} \end{cases}\right] \quad (4)$$

$A_m$, $B_m$, $C_m$, $D_m$, $G_m$, and $J_m$ are unknown amplitudes, and $\alpha_n = \sqrt{k_0^2\varepsilon_n - k_{z'}^{(m)2}}$.

This equation alone has in it 7 unknowns. Fortunately $A_m$ and $B_m$ can be found via the upper and lower boundary conditions, but $C_m$, $D_m$, $G_m$, $J_m$, and $k_{z'}^{(m)}$ must be solved for. To do this we must use Maxwell's equations and the boundary conditions for $E$ and $H$ fields at metal/dielectric boundaries. For *p* polarization ($\boldsymbol{F} \to \boldsymbol{H}$):

$$E_x = \frac{k_{z'}^{(m)}}{k_0\varepsilon}H_{y'} \quad (5)$$

$$E_{z'} = \frac{i}{k_0\varepsilon}\frac{\partial H_{y'}}{\partial x} \quad (6)$$

The field components $H_{y'}$ and $E_{z'}$ are continuous across the boundary therefore:

$$C_m e^{i\alpha_1 d_1} + D_m e^{-i\alpha_1 d_1} = G_m + J_m \quad (7)$$

$$\frac{\alpha_1}{k_0\varepsilon_1}(C_m e^{i\alpha_1 d_1} - D_m e^{-i\alpha_1 d_1}) = \frac{\alpha_2}{k_0\varepsilon_2}(G_m - J_m) \quad (8)$$

Here $d_1$ is the width of the first layer. In matching the period boundary we must take into account the phase factor $e^{ik_x d}$ in order to be able to match the phase in layers 1 and 3 to this one. This leads to the equations

$$(C_m + D_m)e^{ik_x d} = G_m e^{i\alpha_2 d_2} + J_m e^{-i\alpha_2 d_2} \quad (9)$$

$$\frac{\alpha_1}{k_0\varepsilon_1}(C_m - D_m)e^{ik_x d} = \frac{\alpha_2}{k_0\varepsilon_2}(G_m e^{i\alpha_2 d_2} - J_m e^{-i\alpha_2 d_2}) \quad (10)$$

Where $d_2$ is the width of the second layer.

These four boundary conditions result in the Kronig-Penny (KP) equation:

$$\cos\alpha_1 d_1 \cos\alpha_2 d_2 - \frac{1}{2}\left(\frac{p_1}{p_2} + \frac{p_2}{p_1}\right)\sin\alpha_1 d_1 \sin\alpha_2 d_2 = \cos k_x d \quad (11)$$

with $p_i = \frac{\alpha_i}{k_0\varepsilon_i}$. For *s* polarization ($\boldsymbol{F} \to \boldsymbol{E}$) the characteristic equation is similar, except that $p_i = -\frac{\alpha_i}{k_0}$.

The KP equation provides the means to find $k_{z'}^{(m)}$ and the corresponding $C_m$, $D_m$, $G_m$, and $J_m$ coefficients. Unfortunately, the KP equation is transcendental and has an infinite number of roots. It is however quite possible to find a finite set of roots for an individual set of parameters [8, 9]. But to get any directly relatable data we need to be able to look at a wide swath of the parameter space with good resolution simultaneously.

The process to find roots for a single set of parameters is to first choose a maximum value for $\left|k_{z'}^{(m)}\right|$, this gives a minimum decay length and wavelength to be considered. Then we create a graph overlaying the zero contours of the real and imaginary parts of the KP equation as a function of the real and imaginary parts of $k_{z'}^{(m)2}$ (see Fig. 2). The desired roots are at the intersections of these contours.

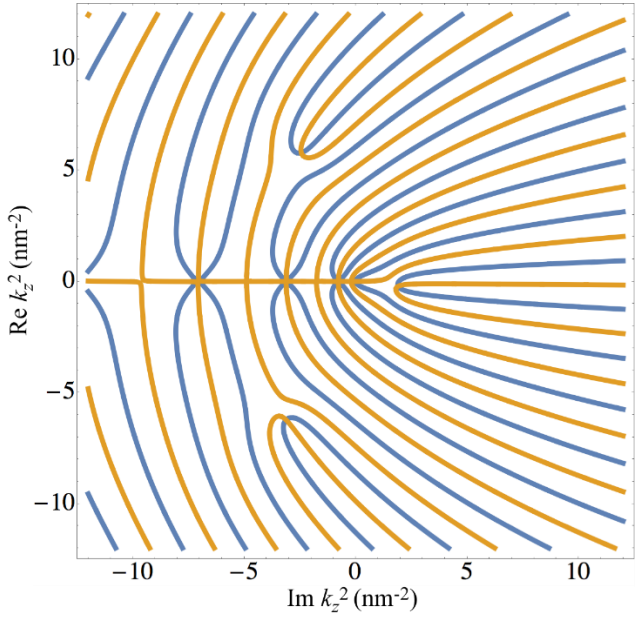

**Figure 2:** A graph of the zero contour curves of the real (blue) and imaginary (orange) parts of the left side of KP equation [Eq. (11)] using 8.5 nm GaAs and 1.5 nm Ag at $\omega = 2.47$ eV and normal incidence ($k_x = 0$).

To find the $k_{z'}^{(m)}$ in the desired range of parameters we use an iterative method in which we first do the above process for one set of parameters. Then we use those roots as the starting point for very slightly different parameters iterating until the whole parameter space is covered.

This process leads to roots jumping from one branch to another even as we reached the upper limit of a reasonable number of iterations. To avoid this we use a pair of equations that split the KP equation into even and odd roots as long as the angle of incidence is zero, in other words $k_x = 0$. [9]

$$p_1 \tan\left(\frac{p_2 d_2}{2}\right) + p_2 \tan\left(\frac{p_1 d_1}{2}\right) = 0 \quad (12)$$

$$p_2 \tan\left(\frac{p_2 d_2}{2}\right) + p_1 \tan\left(\frac{p_1 d_1}{2}\right) = 0 \quad (13)$$

These equations split all the troublesome roots apart into the two separate equations as visible in Fig. 3. However, there are still two roots of Eqn. (13) in $p$ polarization that continue to have this issue. We have gotten around this issue by simultaneously changing multiple parameters in a single step such that the roots change much slower throughout the sections where the roots would normally need much higher resolution.

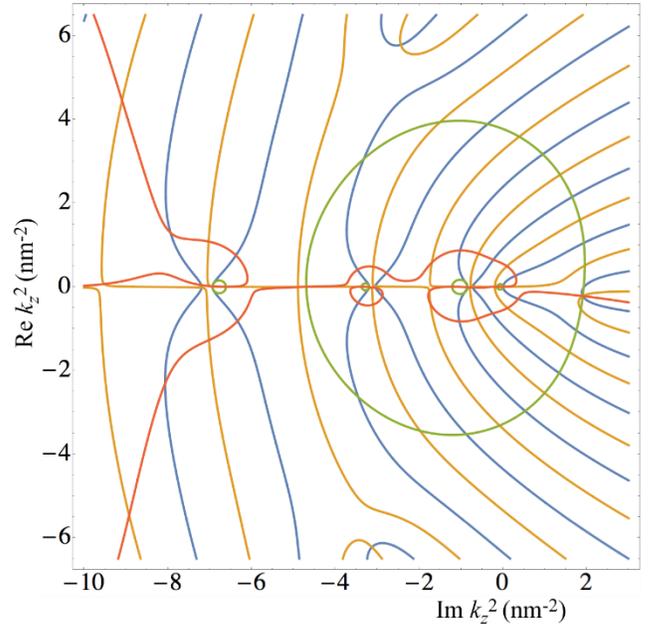

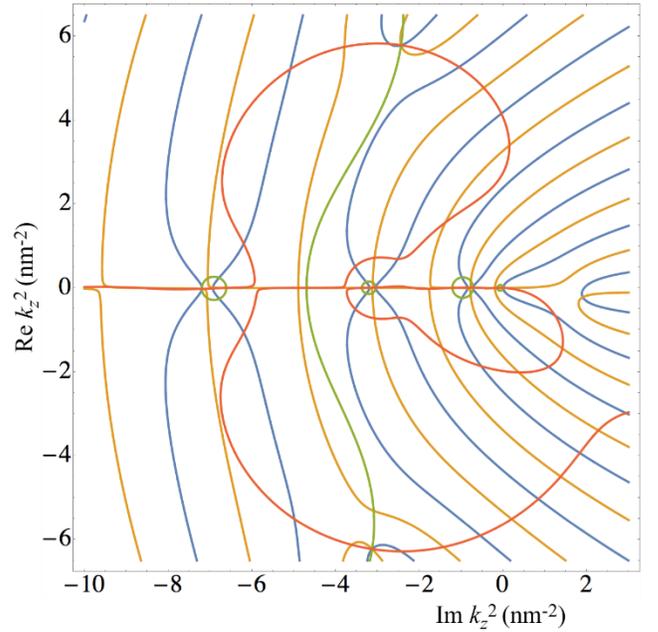

**Figure 3:** A graph of the zero contour curves of the real (green) and imaginary (red) parts of the left sides of Eqs. (12) (top) and (13) (bottom) using the same parameters as Fig. 2 and overlaid on top of Fig. 2.

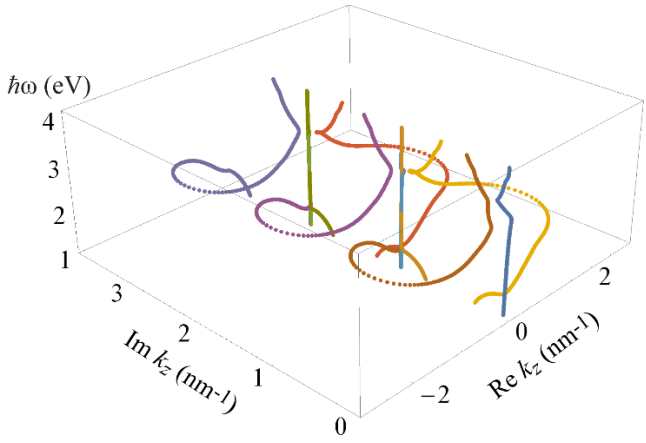

**Figure 4:** Roots of the KP equation [Eq. (11)-(13)] as a function of frequency for 7.5 nm GaAs & 2.5 nm Ag, at normal incidence.

After the modes in layer 2 are found (see Fig. 4) the fields need to be matched at the upper and lower boundaries. First we need to identify all the independent waves that are present.

**Table 1: The waves and their amplitudes within each layer**

| Layer 1 | |
|---|---|
| TM Incident | $P$ |
| TE Incident | $S$ |
| TM Diffraction | $R_{Mn}$ |
| TE Diffraction | $R_{En}$ |
| Layer 2 | |
| p Waveguide | $A_{pm}$ & $B_{pm}$ |
| s Waveguide | $A_{sm}$ & $B_{sm}$ |
| Layer 3 | |
| TM Diffraction | $T_{Mn}$ |
| TE Diffraction | $T_{En}$ |

Amplitudes $P$ and $S$ can be set as desired but all the rest must be found. We follow a usual method for matching infinite sets of plane and waveguide waves [10]. First we set the fields we intend to match equal to each other and multiply through by $e^{-ik_x^{(l)}x}$. Then we integrate both sides over a single period of the structure. This gives

$$\int_0^d e^{ik_x^{(n)}x} e^{-ik_x^{(l)}x} dx = \delta_{nl} d \qquad (14)$$

on the side of layers 1 or 3. As for layer 2 there are terms of the form:

$$\int_0^d e^{-ik_x^{(l)}x} \begin{cases} C_m e^{i\alpha_1 x} + D_m e^{-i\alpha_1 x} & x < d_1 \\ G_m e^{i\alpha_2 x} + J_m e^{-i\alpha_2 x} & x > d_1 \end{cases} dx \qquad (15)$$

At this point we reduced the system to a set of eight matrix equations, one for each $x$ and $y$ component of the $E$ and $H$ fields on the upper and lower boundaries. Then using block matrices we reduce those eight equations to these four:

$$\widehat{X}^{Ax} A + \widehat{X}^{Bx} B + \widehat{K}^{Rx} R = D^x \qquad (16)$$

$$\widehat{X}^y (A + B) + \widehat{K}^{Ry} R = D^y \qquad (17)$$

$$\widehat{W}^{Ax} A + \widehat{W}^{Bx} B + \widehat{K}^{Tx} T = 0 \qquad (18)$$

$$\widehat{W}^{Ay} A + \widehat{W}^{By} B + \widehat{K}^{Ty} T = 0 \qquad (19)$$

Where the $\widehat{X}$s and $\widehat{W}$s contain matrices with entries similar to Eqn. (14) while the $\widehat{K}$s are 2x2 block matrices of diagonal matrices containing coefficients due to angles, derivatives, and the like.

These four equations can be reduced further to a single equation:

$$\widehat{M} V = D \qquad (20)$$

$$\widehat{M} = \begin{pmatrix} \widehat{X}^{Ax} & \widehat{X}^{Bx} & \widehat{K}^{Rx} & \widehat{0} \\ \widehat{X}^y & \widehat{X}^y & \widehat{K}^{Ry} & \widehat{0} \\ \widehat{W}^{Ax} & \widehat{W}^{Bx} & \widehat{0} & \widehat{K}^{Tx} \\ \widehat{W}^{Ay} & \widehat{W}^{By} & \widehat{0} & \widehat{K}^{Ty} \end{pmatrix} \qquad (20a)$$

$$V = \begin{pmatrix} A \\ B \\ R \\ T \end{pmatrix} \& D = \begin{pmatrix} D^x \\ D^y \\ 0 \\ 0 \end{pmatrix} \qquad (20c/d)$$

At this point it seems to be a simple task to invert $\widehat{M}$ to solve for $V$, however when using any roots that decay significantly in layer 2, $\widehat{M}$ quickly becomes so poorly conditioned that even double precision isn't enough to produce anything but zeros (see Fig. 5 and discussion after Eqn. (35)). The major issue is the matrix $\widehat{H}$ and its inverse contained within the $\widehat{W}$s, where $H_{ml} = e^{ik_z^{(m)}h} \delta_{ml}$ and $h$ is the height of layer 2. This issue can be resolved by using the transfer matrix method we developed.

In this method, we consider each boundary independently to find how an incident wave is converted into outgoing waves. Then we propagate and feed the outgoing waves as incident onto the other boundary and so on, which forms an iterative process.

The full set of waves coming off the upper and lower boundaries can be found by constructing the formulas:

$$R = RI + \widehat{R}\widehat{P} B \qquad (21)$$

$$A = AI + \widehat{A}\widehat{P} B \qquad (22)$$

$$B = \widehat{B}\widehat{P} A \qquad (23)$$

$$T = \widehat{T}\widehat{P} A \qquad (24)$$

Here $RI$ ($AI$) is a vector containing the amplitudes of diffraction (waveguide) waves created as a direct result of the incident waves coming from the top medium. $\widehat{A}$ and $\widehat{B}$ are matrices that convert a waveguide wave amplitude vector into a counter-propagating waveguide wave amplitude vector on the upper and lower boundaries respectively. $\widehat{R}$ and $\widehat{T}$ convert waveguide wave amplitude vectors into reflected and transmitted wave amplitude vectors respectively. $\widehat{H}$ is used to propagate the waveguide vectors down or up the structure.

Substituting $B$ into the formula for $A$ we find:

$$A = AI + \widehat{A}\widehat{P}\widehat{B}\widehat{P} A = AI + \widehat{RT} A \qquad (25)$$

Solving for $A$:

$$A = (\hat{1} - \widehat{RT})^{-1} AI \quad (26)$$

Then to avoid taking the inverse we expand Eq. (26) to finally get to the equation:

$$A = (\hat{1} + \widehat{RT} + \widehat{RT}^2 + \widehat{RT}^3 + \cdots) AI \quad (27)$$

Here $\widehat{RT}$ is a matrix, which we call a *round-trip matrix*. It propagates a set of modes at the top boundary to the bottom of layer 2, reflects them, propagates them back and reflects them once more. The expansion (27) can be understood as a sum of a series of roundtrips and eliminates the evanescent modes as they decay. This is the root of the effectiveness of the method.

To use this method we must start by finding **RI** and **AI**. Consider a two-layer system consisting of layers 1 and 2 only. In this case, we deal with P, S, R, and A coefficients. We again matched the x and y components of the E and H fields on the boundary by multiplying by $e^{-ik_x^{(l)}x}$ and integrating. This time having just the four boundary conditions we only get two equations on the first block matrix system:

$$\hat{\chi}^x AI + \widehat{K}^{Rx} RI = D^x \quad (28)$$

$$\hat{\chi}^y AI + \widehat{K}^{Ry} RI = D^y \quad (29)$$

Thus we recreate Eqn. (20) as a 2x2 system making the inversion even simpler, and without a $\widehat{W}$ there is no $\widehat{H}$ and thus $\widehat{M}$ is not poorly conditioned.

The next step is to find $\hat{B}$ and $\hat{T}$. Consider the boundary between layers 2 and 3. In this case we have a set of incident waveguide waves, A, reflected waveguide waves, B, and transmitted diffraction waves, T. Using these waves the system only changes slightly to become:

$$\hat{\chi}^x B + \widehat{K}^{Tx} T = \widehat{D}^{Ax} A \quad (30)$$

$$\hat{\chi}^y B + \widehat{K}^{Ty} T = \widehat{D}^{Ay} A \quad (31)$$

The As will be defined later, the $\widehat{D}$s are constructed in the same way as the $\hat{\chi}$s, and the $\widehat{K}$s are the same sans slight differences due to material choice on the upper and lower boundaries.

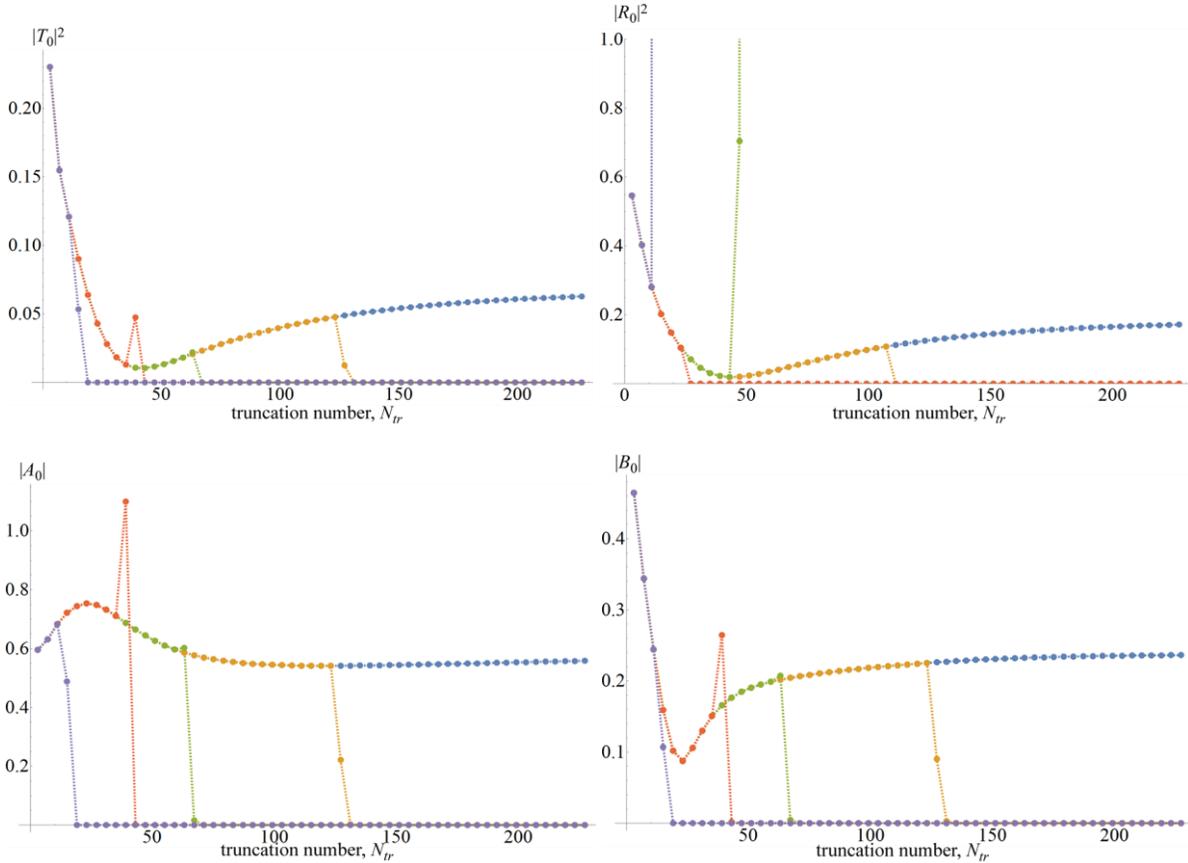

**Figure 5: a comparison of the results of the two different methods for a structure composed of a 100 nm thick Ag (100 nm)/vacuum (100 nm) array on top of a Ag substrate. Blue: transfer matrix method using double precision. Purple: characteristic matrix approach using double precision. Red: characteristic matrix with 32 digit precision. Green: characteristic matrix with 64 digit precision. Orange: characteristic matrix with 128 digit precision. (Non-standard precision done using Mathematica's variable precision)**

$$\widehat{M}^{-1} = \begin{pmatrix} \left(\hat{\chi}^x - \widehat{K}^x \widehat{K}^{y-1} \hat{\chi}^y\right)^{-1} & -\hat{\chi}^{x-1} \widehat{K}^x \left(\widehat{K}^y - \hat{\chi}^y \hat{\chi}^{x-1} \widehat{K}^x\right)^{-1} \\ -\widehat{K}^{y-1} \chi^y \left(\hat{\chi}^x - \widehat{K}^x \widehat{K}^{y-1} \hat{\chi}^y\right)^{-1} & \left(\widehat{K}^y - \hat{\chi}^y \hat{\chi}^{x-1} \widehat{K}^x\right)^{-1} \end{pmatrix} \quad (32)$$

Next, we returned to the boundary between layers 1 and 2. This time the incident waves being waveguide waves. The resultant equations are the same as Eqns. (30)-(31) except $B \to A$, $A \to B$, and $T \to R$.

Now we need to use Eqns. (28)-(31) to create matrices that directly convert a set of incident waves to reflected and transmitted waves.

$$\widehat{M}^{-1} D = V = \begin{pmatrix} AI \\ RI \end{pmatrix} \quad (33)$$

$$\widehat{M} \begin{pmatrix} B \\ T \end{pmatrix} = \begin{pmatrix} \widehat{D}^{Ax} \\ \widehat{D}^{Ay} \end{pmatrix} A \quad (34)$$

$$\therefore$$

$$\widehat{M}^{-1} \begin{pmatrix} \widehat{D}^{Ax} \\ \widehat{D}^{Ay} \end{pmatrix} = \begin{pmatrix} \widehat{B} \\ \widehat{T} \end{pmatrix} \quad (35)$$

To do this we need the $\widehat{M}^{-1}$ associated with each set of equations, the general form being Eqn. (32). Doing Eqns. (34) and (35) again for the other side, mutatis mutandis, yields the $\widehat{A}$ and $\widehat{R}$ conversion matrices.

## 4. RESULTS AND CONCLUSIONS

To compare the boundary condition approach and the transfer matrix method we devised, we plot the magnitudes of amplitudes of 0$^{\text{th}}$ order waves in the structure in Fig. 5. In each panel the blue dots represent the results of the transfer matrix method in double precision. For convergence one needs to truncate the infinite system of equations at about $|k_z|^2 \approx (1\,\text{Å})^{-2}$ which occurs at $N_{tr} \approx 200$ waveguide modes in Fig. 5's structure. This means that taking just one propagating waveguide mode [9] is not enough for determining the optical properties of such structures. The evanescent modes may not contribute to the determination of the spectral positions of the resonances in subwavelength structures, but they determine the power distribution at interfaces between layers. Additionally, the possibility of including multiple modes in our method allows for consideration of large period structures and the establishing of the exact conditions under which the metamaterial approximations fail.

The results of the calculation using the boundary condition method are shown for double precision (purple), 32-digit precision (red), 64-digit precision (green) and 128-digit precision (orange). Even when using 128-digit precision, which requires significantly more time to calculate, the boundary condition method is not capable of reaching the convergence requirement for the system due to poorly conditioned matrices.

Currently, there is a strong interest in applicability of effective medium approximation for describing fields in metamaterials. To evaluate the correctness of the results of this approximation, one has to compare it to an exact calculation. In Fig. 6 we provide a calculation of the total reflectivity of a metal-dielectric array with period of 10 nm, different metal fractions $f$ and suspended in vacuum.

The graph shows a set of alternating Fabri-Perot resonances with special properties, which allow for the polarization rotation effect to be explained more fully in our upcoming paper [11, 12].

To conclude, we have developed a new transfer matrix method for calculating the fields in metal-dielectric parallel-plate arrays. This method allows reaching convergence and obtaining reliable results. We apply this method to model metasurface polarization rotators.

## 1. ACKNOWLEDGEMENTS

M. L. was supported through the Blue Waters Internship program, which included a two-week workshop about MPI, OpenMP, CUDA, and OpenACC, as well as many other useful and interesting topics such as CPU structure at University of Illinois at Urbana-Champaign. M. D. acknowledges the support from the Office of the Vice President for Research & Economic Development and the Jack N. Averitt College of Graduate Studies at Georgia Southern University.

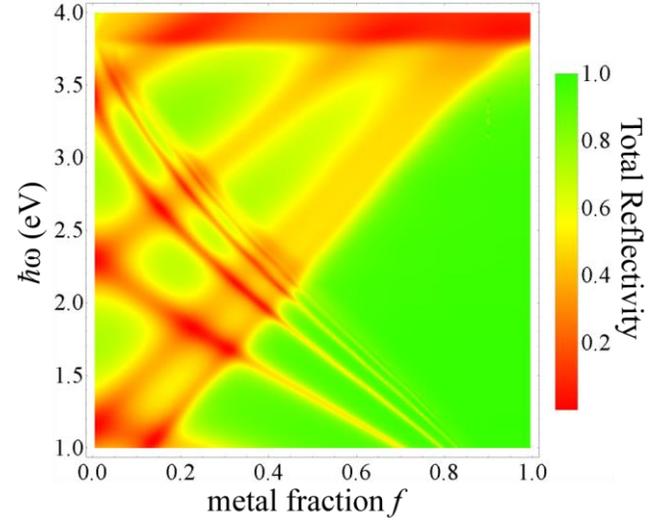

**Figure 6: The reflectivity of a 10nm period, 150nm height structure suspended in air, made of Ag and GaAs smoothly varying from all GaAs (left), to all Ag (right) over varied frequency.**

## 2. REFERENCES

[1] D. Y. K. Ko and J. R. Sambles. Scattering Matrix Method for Propagation of Radiation in Stratified Media: Attenuated Total Reflection Studies of Liquid Crystals. *JOSA A*, 5(11):1863-1866, 1988.

[2] N. P. K. Cotter, T. W. Preist, and J. R. Sambles. Scattering-Matrix Approach to Multilayer Diffraction. *JOSA A*, 12(5):1097-1103, 1995.

[3] V. G. Veselago. The electrodynamics of substances with simultaneously negative values of ε and μ. *Sov. Phys. Usp.*, 10(4):509–14, 1968.

[4] J. B. Pendry. Negative Refraction Makes a Perfect Lens. *Phys. Rev. Lett.*, 85(18):3966–9, 2000.

[5] D. Schurig, et al. Metamaterial Electromagnetic Cloak at Microwave Frequencies. *Science,* 314(5801):977–980, 2006.

[6] Nanfang Yu, Patrice Genevet, Mikhail A. Kats, Francesco Aieta, Jean-Philippe Tetienne, Federico Capasso, and Zeno Gaburro. Light Propagation with Phase Discontinuities: Generalized Laws of Reflection and Refraction. *Science,* 334(6054):333-337, 2011.

[7] D. Keene and M. Durach. Hyperbolic resonances of metasurface cavities. *Opt. Express*, 23(14):18577-88, 2015.

[8] Alexey Orlov, Ivan Iorsh, Pavel Belov, and Yuri Kivshar. Complex band structure of nanostructured metal-dielectric metamaterials. *Opt. Express*, 21(2): 1593-8, 2013

[9] B. Sturman, E. Podivilov, and M. Gorkunov. Theory of Extraordinary Light Transmission through Arrays of Subwavelength Slits. *Phys. Rev. B*, 77(7):075106, 2008.

[10] Ping Sheng, R. S. Stepleman, and P. N. Sanda. Exact eigenfunctions for square-wave gratings: Application to diffraction and surface-plasmon calculations. Phys. Rev. B, 26(6):2907, 1982.

[11] D. Keene, M. Lepain, and M. Durach. Anisotropic Fabry Perot resonances in Metal-Dielectric Meta-Nano-Layer. APS March Meeting Abstracts, vol. 1: 6010, 2015

[12] M. LePain, D. Keene, and M. Durach. Ultrathin Metasurface-based Polarization Rotators. In preparation